\documentclass[aps,prA,twocolumn,superscriptaddress]{revtex4-2}

\usepackage{amsmath}
\usepackage{amsthm}
\usepackage{amssymb}
\usepackage{graphicx}
\usepackage{esint}
\usepackage{titlesec}
\usepackage{amsfonts}
\usepackage{xcolor}
\usepackage{times}
\usepackage[utf8]{inputenc}
\usepackage{amsfonts}
\usepackage{amsmath}
\usepackage{color}
\usepackage{footmisc}
\usepackage{bbold}
\usepackage{amsthm}
\usepackage{graphicx}
\usepackage{halloweenmath} 
\usepackage{curve2e} 
\usepackage{titlesec}
\usepackage{hyperref}

\hypersetup{
    colorlinks=true,
    linkcolor=black,
    citecolor = black,
    filecolor=magenta,      
    urlcolor=cyan
    }

\titleformat{\subsection}[runin]
{\normalfont\bfseries}
{\thesubsection}{.5em}{}[  ~~]
\titlespacing{\subsection}
{\parindent}{1.5ex plus .1ex minus .2ex}{0pt}

\titleformat{\subsubsection}[runin]
{\it}
{\S\ \thesection.}{.5em}{}[  ~~]
\titlespacing{\subsubsection}
{\parindent}{1.5ex plus .1ex minus .2ex}{0pt}

\makeatletter

\@ifdefinable\SuCmathpictvertex{} 
\@ifdefinable\@SuC@reserved@dimen{\newdimen\@SuC@reserved@dimen}

\newenvironment*{@SuC@math@picture}[8]{%
  \def\SuCmathpictvertex{\circle*{#6}}%
  \setlength\unitlength{\fontdimen 22 #5\tw@}%
  \setlength\@SuC@reserved@dimen{#7\unitlength}%
  \kern\@SuC@reserved@dimen
  \@HwM@d@pict@strut{#2}%
  \picture(#3,#1)(#4,-1)%
    \roundcap
    \roundjoin
    \linethickness{#8\@HwM@thickness@units@for #5}%
}{%
  \endpicture
  \kern\@SuC@reserved@dimen
}
\newcommand*\@SuC@general@pict[9]{%
  \begin{@SuC@math@picture}%
            {#2}{#3}
            {#4}{#5}
            #6
            {#7}
            {#8}
            {#9}
    #1%
  \end{@SuC@math@picture}%
}
\newcommand*\@SuC@math@version@shunt[7]{%
  \@HwM@choose@thicknesses{\@SuC@general@pict {#1}{#2}{#3}{#4}{#5}#7}%
      {{.4}{.2}{}}
      {{0.5}{.25}{0.75}}
}

\newcommand*\DeclareNewSuCMathPict[6]{%
  \newcommand*{#1}{%
    \@HwM@general@ordinary@symbol
      {\@SuC@math@version@shunt {#6}{#2}{#3}{#4}{#5}}%
  }%
}

\makeatother

\DeclareNewSuCMathPict{\pentagon}
            {3}{1}  
            {4}{-2} 
{
    \polygon(90:2)(162:2)(234:2)(306:2)(378:2)%
    \put (90:2){\SuCmathpictvertex}%
    \put(162:2){\SuCmathpictvertex}%
    \put(234:2){\SuCmathpictvertex}%
    \put(306:2){\SuCmathpictvertex}%
    \put(378:2){\SuCmathpictvertex}%
}

\newcommand{\vect}[1]{\vec{#1}}
\newcommand{\ket}[1]{\left| {#1}\right\rangle}

\newcommand{\braket}[2]{\left\langle {#1}\middle|{#2} \right\rangle}
\newcommand{\braAket}[3]{\left\langle {#1}\middle|{#2}\middle|{#3} \right\rangle}
\newcommand{\op}[1]{\hat{#1}}
\newcommand{\perm}[1]{\text{perm}\left({#1}\right)}

\theoremstyle{definition}
\newtheorem{conjecture}{Conjecture}
\newtheorem*{counterexample*}{Counter-example}

\newcommand{\ad}{\hat{a}^{\dagger}}

\newcommand{\Bd}{\hat{B}^{\dagger}}
\newcommand{\cd}{\hat{c}^{\dagger}}
\newcommand{\cdprime}{\hat{c'}^{\dagger}}

\newcommand{\ahd}[1]{\hat{a}^{\dagger}_{h, {#1}}}
\newcommand{\avd}[1]{\hat{a}^{\dagger}_{v, {#1}}}

\newcommand{\Pdist}[1]{P_{#1}^{(\text{dist.})}}
\newcommand{\Pbos}[1]{P_{#1}^{(\text{bos})}}
\newcommand{\Pferm}[1]{P_{#1}^{(\text{ferm})}}

\newcommand{\valery}{Shchesnovich}
\newcommand{\Stranspose}{S^{T}}

\begin{document}

\title{Boson bunching is not maximized by indistinguishable particles}

\author{Beno\^it Seron}
\email{benoitseron@gmail.com}
\affiliation{Quantum Information and Communication, Ecole polytechnique de Bruxelles, CP 165/59, Universit\'e libre de Bruxelles, 1050 Brussels, Belgium}

\author{Leonardo Novo}
\email{lfgnovo@gmail.com}
\affiliation{Quantum Information and Communication, Ecole polytechnique de Bruxelles, CP 165/59, Universit\'e libre de Bruxelles, 1050 Brussels, Belgium}

\author{Nicolas J. Cerf}
\email{ncerf@ulb.ac.be}
\affiliation{Quantum Information and Communication, Ecole polytechnique de Bruxelles, CP 165/59, Universit\'e libre de Bruxelles, 1050 Brussels, Belgium}

\begin{abstract}
Boson bunching is amongst the most remarkable features of quantum physics. A celebrated example in optics is the Hong-Ou-Mandel effect, where the bunching of two photons arises from a destructive quantum interference between the trajectories where they both either cross a beam splitter or are reflected. This effect takes its roots in the indistinguishability of identical photons. Hence, it is generally admitted -- and experimentally verified -- that bunching vanishes as soon as photons can be distinguished, e.g., when they occupy distinct time bins or have different polarizations. Here we disproof this alleged straightforward link between indistinguishability and bunching by exploiting a recent finding in the theory of matrix permanents. We exhibit a family of optical circuits where the bunching of photons into two modes can be significantly boosted by making them partially distinguishable via an appropriate polarization pattern. This boosting effect is already visible in a 7-photon interferometric process, making the observation of this phenomenon within reach of current photonic technology. This unexpected behavior questions our understanding of multiparticle interference in the grey zone between indistinguishable bosons and classical particles.

\end{abstract}

\maketitle

In quantum physics, it is common knowledge that a gain of information results in the extinction of quantum interference. In the iconic double-slit experiment, the interference fringes originate from the absence of which-path information \cite{ wootters1979complementarity, greenberger1988simultaneous, mandel1991coherence}. As emphasized for instance by Feynman \cite{feynman1963lectures}, the fringes necessarily disappear as soon as the experiment allows us to learn that the particles have taken one or the other path. In quantum optics, photon bunching is another distinctive feature that follows from quantum interference. 
Specifically, the indistinguishability of photons makes it such that one cannot know which photon has followed a given trajectory in a linear interferometer (see Fig.~\ref{fig:generalInterferometer}). The Hong-Ou-Mandel (HOM) effect \cite{hong1987measurement}, for example, arises from the fact that one cannot distinguish the trajectory where two photons have crossed a 50:50 beam splitter from the trajectory where they were both reflected. The net result is a tendency of indistinguishable photons to occupy the same mode -- that is, to bunch -- as a consequence of this lack of information. In accordance with Feynman's rule of thumb, quantum interference effects become less pronounced as soon as the photons become distinguishable, for example if they occupy distinct temporal or polarization modes so that one gains information about their individual trajectories. For example, the HOM dip disappears for two photons with orthogonal polarization since their distinct trajectories can then be fully traced back. Hence, it is commonly admitted that, even in more general scenarios involving larger interferometers, bunching effects are maximum for fully indistinguishable photons and gradually decline when photons are made increasingly distinguishable \cite{general_rules_bunching, shchesnovich2016universality, carolan2014experimental}.

Here, we find a quantum interferometric scenario that goes against this intuition and contradicts the very idea that distinguishability undermines photon bunching. We consider the probability of multimode bunching, i.e., the probability that all photons entering a linear interferometer end up in a certain subset of the output modes.
In accordance with a longstanding mathematical conjecture on matrix permanents due to Bapat and Sunder \cite{bapat1985majorization}, this bunching probability must indeed be maximum if the input state consists of fully indistinguishable photons \cite{shchesnovich2016universality}. However, inspired by a counter-example to this conjecture recently discovered by Drury \cite{drury2016counterexample}, we have found that multimode bunching may, against all odds, be enhanced if photons are partially distinguishable. Specifically, we construct a family of interferometers such that a higher two-mode bunching probability is attained if the photons are prepared in a well-chosen polarization state (making them partially distinguishable) rather than in the same state (in which case they would all be indistinguishable). In other words, gaining partial information about the photons paradoxically results in a higher probability for them to coalesce on two output modes, contradicting common knowledge. We give an interpretation of the physical process behind this counterintuitive effect and prove that, in our setup, the violation of the two-mode bunching probability grows at least linearly with the number of photons. In the simplest case, an enhancement of 7\% is already visible for 7\,photons in 7\,modes with a specific polarization pattern, which makes the observation of this remarkable phenomenon within reach of today's photonic technology. 

\vspace{7pt}


\section*{Multimode  bosonic bunching}

\noindent Consider a general interferometric experiment in which $n$ bosons are sent through a linear interferometer $\hat{U}$ of $m$ modes, described by the $m\times m$ unitary matrix $U$. Although our discussion is valid for any bosonic particle, we focus on photons here since, in practice, controlled linear interferometric experiments are easier to carry out in photonics. In Ref.~\cite{shchesnovich2016universality}, Shchesnovich provides compelling evidence for the following conjecture: 

\begin{conjecture}[Generalized Bunching]
\label{conj:generalizedBunching}
Consider any input state of classically correlated photons. For any linear interferometer $\hat{U}$ and any nontrivial subset $\mathcal{K}$ of output modes, the probability that all photons are found in $\mathcal{K}$ is maximal if the photons are (perfectly) indistinguishable. 
\end{conjecture}

\begin{figure}[t]
    \centering
    \includegraphics[width = 0.4\textwidth]{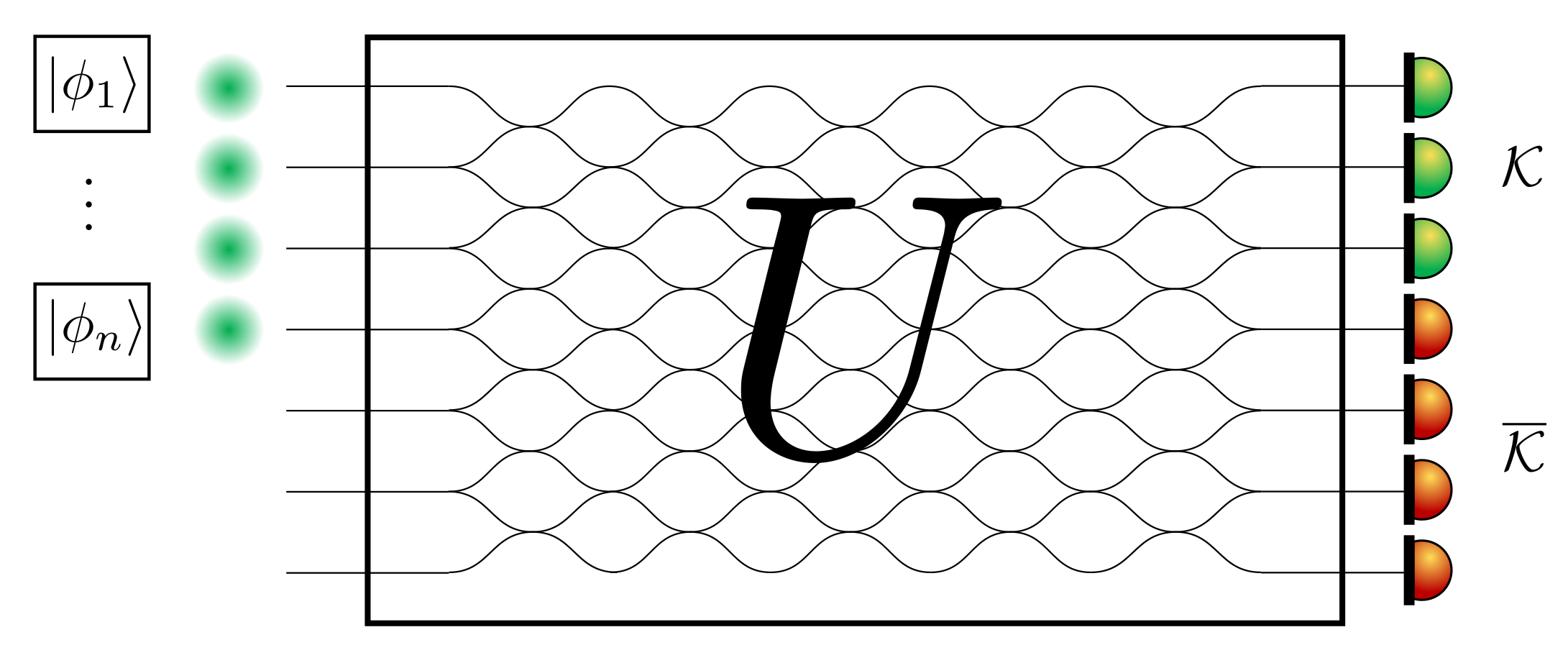}
    \caption{\textbf{Interferometric setup.} Single photons are sent through the first $n$ input modes of a $m$-mode linear interferometer $U$, which can always be decomposed into a network of two-mode couplers~\cite{Clements}. Each photon at input $j$ carries internal degrees of freedom (polarization, arrival time, etc.) described by an (internal) wave function $\ket{\phi_j}$. Not perfectly overlapping wave functions (measured via the Gram matrix $S$) give rise to partial distinguishability amongst the photons, reducing the degree of quantum interference and bosonic effects such as bunching (see Conjecture \ref{conj:generalizedBunching}). We focus in particular on the probability that all $n$ photons bunch in a subset $\mathcal{K}$ (corresponding to the green detectors) of the output modes, while the red detectors in $\overline{ \mathcal{K}}$ do not click. This is a natural extension of the HOM experiment for more than two modes.}
    \label{fig:generalInterferometer}
\end{figure}

 To simplify the mathematical treatment of the problem, we assume here that each photon is prepared independently of each other, i.e., they are uncorrelated, and that there is only one photon in each of the first $n$ modes. Moreover, we also assume that the state of each photon is pure. This setting, depicted in Fig.~\ref{fig:generalInterferometer}, is enough to demonstrate that Conjecture~\ref{conj:generalizedBunching} is false. We refer to Ref.~\cite{shchesnovich2016universality} for a mathematical treatment of bunching probabilities in more general scenarios. 

The state of each photon is not only described by the spatial mode it occupies but also by other degrees of freedom, such as its polarization and its spectral distribution, which we will refer to as internal degrees of freedom. Partial distinguishability can then be modelled by considering that the internal state of the photon entering mode $j$ is described by an (internal) wavefunction $\ket{\phi_j}$. Let us denote the creation operator associated to this state as $\ad_{j, \phi_j}$. We make the common assumption that the interferometer acts only on spatial modes, leaving the internal state of the photon invariant \cite{tichy2015_partial_distinguishability, shchesnovich2015partial, dittelPRA}. More precisely, the action of $\hat{U}$ is described by the following equation
\begin{equation}\label{eq:assumptionU}
    \hat{U}\ad_{j, \phi_j}\hat{U}^{\dagger}= \sum_k U_{jk}\ad_{k, \phi_j},~~\forall j.
\end{equation}
Following Refs.~\cite{tichy2015_partial_distinguishability, shchesnovich2015partial}, it can be shown that the probabilities of the different outcomes of the linear interferometric process not only depend on the unitary $U$ but also on the \emph{distinguishability matrix}, defined as 
\begin{equation}
S_{ij} = \braket{\phi_i}{ \phi_j}. 
\end{equation}
This is a $n\times n$ Gram matrix constructed from all possible overlaps of the internal wave functions of the input photons. In particular, $S = \mathbb{1}$ if all photons are fully distinguishable, while $S = \mathbb{E}$ (with $\mathbb{E}_{i,j} = 1 $ for all $i,j$) in the case where they are fully indistinguishable \footnote{The internal wavefunction of each photon may be multiplied by a phase $\ket{\phi_i} \!\rightarrow\! e^{i\theta_i}\ket{\phi_i}$, which does not affect event probabilities. Hence, there is an equivalence class of distinguishability matrices $S$ for each physical situation as $S_{ij} \equiv e^{i(\theta_j - \theta_i)}S_{ij}$. Here, we represent the equivalence class of fully indistinguishable particles with a single matrix $S = \mathbb{E}$.\label{footnote:physicalEquivalencyClass}}. Intermediate situations between these two extreme cases are called \emph{partially distinguishable.} 

In order to compute the probability that all $n$ photons are found in a subset $\mathcal{K}$ of the output modes, which we refer to as the {\it multimode bunching probability} $P_n(S)$, it is useful to define the matrix
\begin{equation}
    \label{eq:H_Shchesnovich}
    H_{a,b} = \sum _{l \in \mathcal{K}} U_{l,a}^* U_{l,b},  
\end{equation}
where $a,b \in \{1, ...,n\}$. For a fixed interferometer $U$ and subset $\mathcal{K}$, the multimode bunching probability is a function of the distinguishability matrix $S$ and can be expressed as
\begin{equation}
    \label{eq:bunchingProbability}
    P_n(S)  = \perm{H \odot \Stranspose} 
\end{equation}
i.e., the permanent of the Hadamard (or elementwise) product $(H\odot \Stranspose)_{ij} \equiv H_{ij}S_{ji}$ \cite{shchesnovich2016universality} (see also Appendix \ref{sec:bunchingProbability}). It is important to remark that $H$, $S$ and $H\odot \Stranspose$ are all positive semidefinite matrices, which ensures their permanent is positive \cite{zhang2011matrix}. Moreover, note that for indistinguishable photons,  $\Pbos{n} = \perm{H}$. Hence, Conjecture~\ref{conj:generalizedBunching}, when restricted to the setting that we consider (see Fig.~\ref{fig:generalInterferometer}), takes the following mathematical form: 
\begin{equation}\label{eq:bunching_conjecture}
\perm{H \odot \Stranspose} \stackrel{?}{\leq} \perm{H },  
\end{equation} 
with equality holding if $S$ corresponds to indistinguishable photons. The reasons to presume that  this conjecture might be an actual physical law governing multiparticle interferences are manifold and, in what follows, we detail several evidences supporting this hypothesis. 

\subsubsection*{Single-mode bunching.} If the subset $\mathcal{K}$ is a single output mode, then the conjecture holds. In this case, if we choose $\mathcal{K}=\{1\}$, the single-mode bunching probability is given by
\begin{equation}\label{eq:singlemode_bunch}
P_n(S)= \prod_{i=j}^n |U_{1j}|^2\, \perm{S}= \Pdist{n}\, \perm{S}, 
\end{equation}   
where $\Pdist{n}=P_n(\mathbb{1}) $ corresponds to the multimode bunching probability when the photons are fully distinguishable. Then, $P_n(S)$ is indeed maximum for fully indistinguishable photons since the maximum value of $\perm{S}$ is attained when $S= \mathbb{E}$, with $\perm{\mathbb{E}}=n!$. This is also a reason why $\perm{S}$ can be seen as a measure of indistinguishability of the input photons \cite{tichy2015_partial_distinguishability, shchesnovich2015tight}. Note that the celebrated HOM effect can be recovered from Eq.~\eqref{eq:singlemode_bunch}: for a two-mode 50:50 beam-splitter, the maximum probability of bunching in a single output mode is attained for fully indistinguishable photons and given by $1/4 \times 2! = 1/2$.

\begin{figure*}
    \centering
    \includegraphics[width=0.8\textwidth]{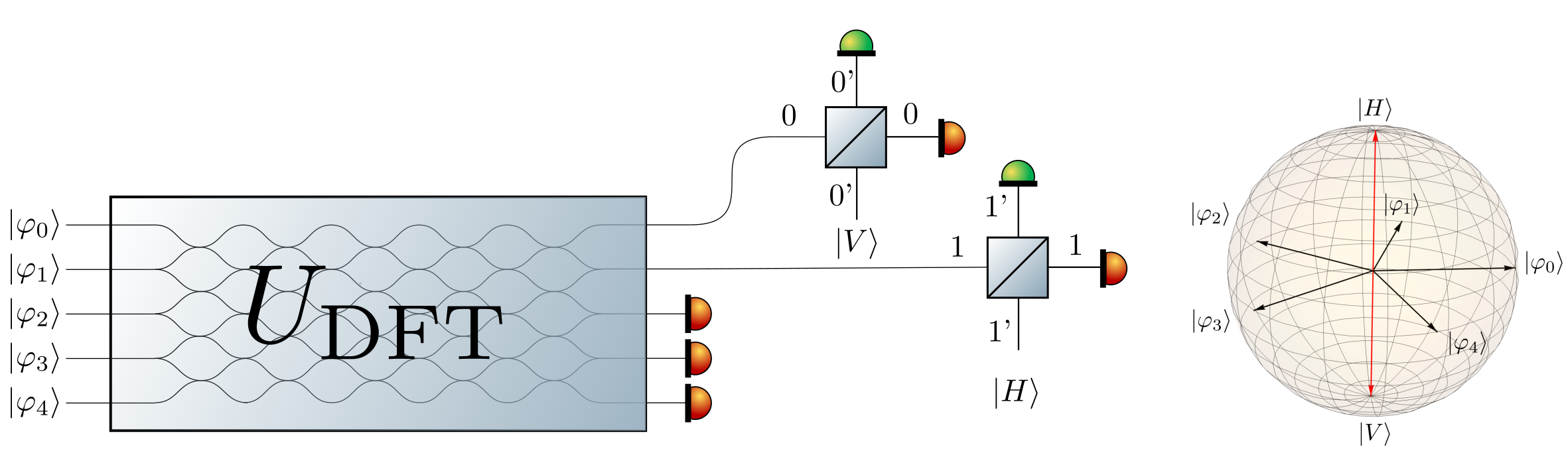}
    \caption{
    \textbf{Boosted two-mode bunching.} (Left) A 7-mode interferometer which violates the generalized bunching conjecture for an appropriate input polarization pattern (depicted in the right panel). Here, 7 photons are sent into the 7 input modes and the probability of detecting them all in the two output modes indicated with green detectors exceeds its value for indistinguishable photons (all with the same polarization). We assume the action of the interferometer is polarization independent, see Eq.~\eqref{eq:assumptionU}. This setup can be generalized to $n$ modes as follows. Defining $q = n-2$, we first send $q$ photons in a $q$-mode Discrete Fourier Transform (DFT) interferometer $U_{jk} = \frac{1}{\sqrt{q}} \omega^{jk}$ with polarization states $\ket{\varphi_j} = \frac{1}{\sqrt{2}}(\ket{H} + \omega^j \ket{V})$, where $j,k = 0,\dots, q-1$ and $\omega = \exp \left(2 i \pi / q\right)$. The upper two output modes (labelled 0 and 1) of the DFT are then sent to two beam splitters of equal transmittance $\eta = 2/n$, achieving interference respectively with a vertically-polarized photon (in mode 0') and horizontally-polarized photon (in mode 1'). We measure the bunching of all $n$ photons in the subset $\mathcal{K}$  corresponding to the  output modes $0'$ and $1'$ indicated with green detectors, thus all red detectors do no click. For $n\ge 7$, we observe a boosted 2-mode bunching probability by comparison with indistinguishable photons. 
    (Right)~Bloch-sphere representation of the input polarization pattern for $n = 7$. The polarization states of the 5 input photons of the DFT (indicated as black arrows) are equally spaced along the equator of the Bloch sphere. We call this special state a 5-star polarization state (or $q$-star polarization state for general $q$) and denote it as $\star$. The two extra photons (indicated as red arrows) have antipodal  -- horizontal and vertical -- polarization states.  
    }
    \label{fig:opticalScheme}
\end{figure*}
In addition, even if we take $|\mathcal{K}| > 1$ but restrict to compare fully indistinguishable with fully distinguishable photons, it appears that the multimode bunching probability satisfies $P_n(\mathbb{E}) \ge P_n(\mathbb{1})$, see  \cite{shchesnovich2016universality}, further suggesting that any partial distinguishability is bound to  decrease bunching effects.

\subsubsection*{Fermion antibunching.} Another physical motivation for Conjecture~\ref{conj:generalizedBunching} is the fact that an analogous statement on fermion antibunching can be proved. Indeed, an input state of $n$ fully indistinguishable fermions \emph{minimizes} the probability that all of the $n$ fermions are bunched in any subset $\mathcal{K}$ of the output modes. This can be shown using a famous result by Schur \cite{schur1918endliche, shchesnovich2016universality}. In the setting we consider, the fermionic multimode bunching probability obeys
\begin{equation}\label{eq:bunching_conjecture_fermions}
\Pferm{n}(S)=\det(H \odot \Stranspose) \geq \det{H },  
\end{equation} 
which follows from the Oppenheim inequality for determinants \cite{oppenheim1930inequalities} stating that for any two positive semidefinite matrices $A$ and $B$ we have
\begin{equation}
    \label{eq:oppenheim}
    \det(A\odot B) \geq \det A \, \det B.
\end{equation}

\subsubsection*{Bapat-Sunder conjecture.} In 1985, Bapat and Sunder have questioned whether an analogue to the Oppenheim inequality holds for permanents \cite{bapat1985majorization}. They conjectured the following:
\begin{conjecture}[Bapat-Sunder]
\label{conj:bapatSunder}
For any two positive semi-definite $n\times  n$ matrices $A = (a_{ij})$ and $B = (b_{ij})$, we have
\begin{equation*}
    \perm{A\odot B} \leq \perm{A} \prod_{i=1}^n b_{ii}.
\end{equation*}
\end{conjecture}

It is easy to see that if this statement was valid, it would imply Eq.~\eqref{eq:bunching_conjecture}. More generally, it would also imply the validity of Conjecture~\ref{conj:generalizedBunching} \cite{shchesnovich2016universality}.

\begin{counterexample*}[Drury]
The Bapat-Sunder conjecture was recently disproved by Drury, who found a 7-dimensional counterexample \cite{drury2016counterexample}. 
It consists in a positive semidefinite matrix $A$ of dimension 7, whose diagonals are $a_{ii}=1$, which is such that 
\begin{equation}\label{eq:violation}
    \frac{\perm{A \odot A^{T}}}{\perm{A}}=\frac{1237}{1152}\approx 1.07 ,
\end{equation}
thus implying a violation of Conjecture \ref{conj:bapatSunder}. For readability, we only give its Cholesky decomposition $A = M^\dagger M$ with 
\begin{equation}
\label{eq:internalWaveFunctions}
M =\frac{1}{\sqrt{2}}\left(\begin{array}{ccccccc}
\sqrt{2} & 0 & 1 & 1 & 1 & 1 & 1 \\ 
 0 & \sqrt{2} & 1 & \omega^{1} & \omega^{2} & \omega^{3} & \omega^{4}
\end{array}\right)
\end{equation}
where $\omega = \exp (2i\pi/5)$ is the fifth root of unity.  As we shall see, this implies that there are 7-dimensional matrices $H$ and $S$ for which Eq.~\eqref{eq:bunching_conjecture} is false, hence contradicting Conjecture \ref{conj:generalizedBunching}.
\end{counterexample*}


Consequently, in spite of the compelling evidences listed above suggesting that multimode bunching should be maximum for indistinguishable bosons, Drury's counterexample allows us to predict the existence of {\it enhanced boson bunching} with partially distinguishable bosons. Before turning to the optical realization and physical mechanism behind this counterintuitive phenomenon, we stress that brute-force numerical trials really seem to support the (now proven wrong) Conjectures \ref{conj:generalizedBunching} and \ref{conj:bapatSunder}. In particular, we generated $10^7$ samples of dimension $n=7$ and of rank $r=2$ with the following physically-inspired procedure: a unitary $U$ is sampled randomly according to the Haar measure and used to compute the matrix $H$. The matrix $S$ is constructed by taking random normalized vectors in a space of dimension $r$. We did not encounter a single counterexample, which seems to imply that Conjecture \ref{conj:generalizedBunching} holds in practically all cases, even when restricting to a dimension and rank where we know that a counterexample actually exists.
Numerical trials of Conjecture \ref{conj:bapatSunder} with two random Gram matrices also gave similar results. These observations are corroborated by the numerical searches reported in Ref.~\cite{shchesnovich2016permanent} and make the finding of enhanced boson bunching via partial distinguishibility even more surprising.
\section*{Optical realization of boosted bunching}
\noindent As we shall see, Drury's counterexample entails the existence of a physical experiment that could violate the generalized bunching conjecture. In Fig.~\ref{fig:opticalScheme} (left panel), we present a possible optical setup realizing this violation and involving \mbox{7-photon} interferometry. 
The details on how to construct the internal states of the photons $\ket{\varphi_j}$ and the unitary matrix $U$
to obtain the desired $H$ and $S$ matrices are explained in Appendix \ref{sec:physicalRealization}. The setting is surprisingly simple: the \mbox{7-mode} interferometer $U$ is constructed with a 5-mode Discrete-Fourier Transform (DFT) supplemented with two additional beam splitters (with the same transmittance $\eta=2/7$). Since $S$ is of rank 2, the internal states live in a two-dimensional Hilbert space, hence it is most natural to use photon polarization, which can easily be manipulated via waveplates (other encodings would also be possible, such as time-bin encoding). 

\begin{figure}
    \centering
    \includegraphics[width = 0.4\textwidth]{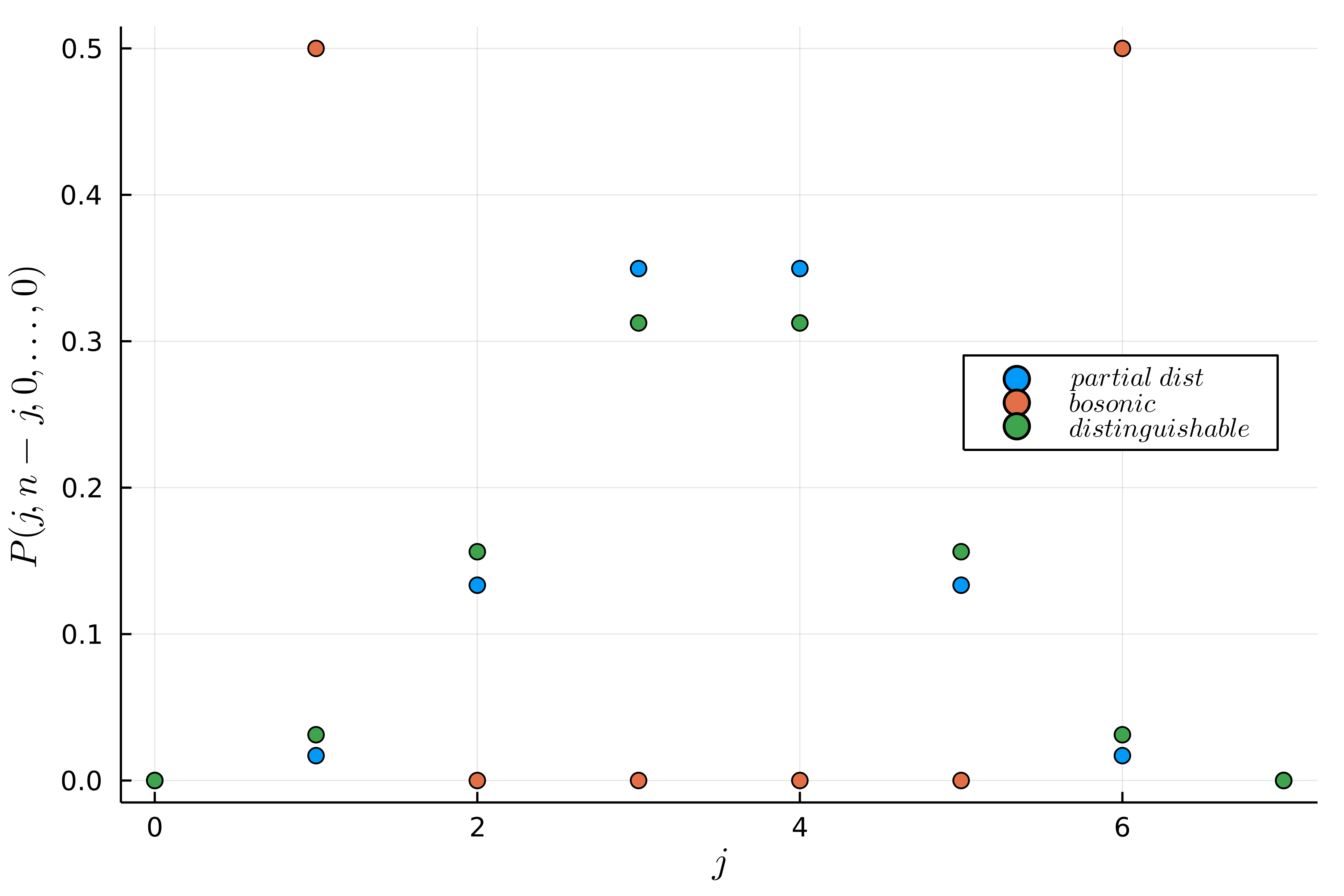}
    \caption{Photon-number probability distributions in mode $0'$ at the output of the 7-mode circuit of Fig.~\ref{fig:opticalScheme} for different scenarios. The distributions are normalized, i.e., the probabilities are conditioned on events where all 7 photons end up in modes 0' and 1' (two-mode bunching events). Due to the symmetry of the circuit, the probability of event $(j,7-j)$ is the same as event $(7-j,j)$. For fully indistinguishable bosons (red points), most events are interferometrically suppressed, originating from the fact that the output of the 5-mode DFT is a NOON state, see Eq.~\eqref{eq:NOON}. The only surviving events are (1,6) and (6,1). Using partially distinguishable bosons (blue points) erases these destructive interferences and, as proven in this work, enhances the overall two-mode bunching probability. This also leads to a qualitatively very different photon-number distribution, which has a Bell-like shape. Interestingly, fully distinguishable particles (green points) lead to a photon-number distribution of similar shape but a significantly smaller two-mode bunching probability. The two-mode bunching probabilities in the three cases are   $\Pbos{7}\approx 7 \times 10^{-3}$, $P_7^{({\displaystyle\star})}\approx 7.5 \times 10^{-3}$ and $P_7^{({\mathrm{dist.}})}\approx 1.5\times 10^{-4}$, respectively. }
    \label{fig:modeDistribution}
\end{figure}

A single photon is sent in each of the 7 input modes with an appropriately chosen polarization state, making the 7 photons  partially distinguishable. The polarization pattern is shown in Fig.~\ref{fig:opticalScheme} (right panel) and can be viewed as a 5-star polarization state denoted as $\star$ (for the 5 photons entering the DFT),     
supplemented with a horizontally-polarized and a vertically-polarized state. The probability of detecting all 7 photons in the output bin (modes labelled by $0'$ and $1'$) is then given by $P_7^{({\displaystyle\star})}\approx 7.5 \times 10^{-3}$. By comparison, for fully indistinguishable photons (all with the same polarization), this probability is only $\Pbos{7}\approx 7 \times 10^{-3}$. This simple experimental setup thus exhibits a  bunching violation greater than $7\%$, in accordance with Eq.~\eqref{eq:violation}. The photon number distribution in modes $0'$ and $1'$ is depicted in Fig.~\ref{fig:modeDistribution} for different scenarios.
%

The outstanding recent progress in boson sampling experiments indicates that the experimental observation of such a boosted bunching due to partial distinguishability should be possible with present-day technology \cite{flamini2018photonic, bosonSamplingRate, hoch2021boson, wang2019boson, wang2020integrated}. For example, 5-photon coincidence rates in the hundreds of Hz range \cite{wang2019boson} have already been demonstrated in optical circuits of a much bigger size that the one represented in Fig.~\ref{fig:opticalScheme}. Moreover, optical implementations of the DFT of dimension 6 and 8 have already been reported \cite{carolan2015universal, crespi2016suppression}. The required photon number resolution (up to 7 photons) could be achieved with single-photon resolving detectors by first multiplexing the output modes $0'$ and $1'$ into several spatial or temporal modes \cite{PNR2019,PNR2020}. As a further feasibility argument, we note that the experimental scheme realizing this boosted bunching is stable under small perturbations to the matrix elements of the unitary $U$ as well as to the distinguishability matrix $S$, see Fig.~\ref{fig:gramPerturbations}. This is easy to understand intuitively as the permanent is a sum of products of matrix elements, thus smooth and well behaved under Taylor expansion (see Appendix \ref{sec:resilience} for a formal proof). A fully realistic treatment of how perturbations affect the bunching violation would depend on the particular details of the physical implementation of the scheme and is out of the scope of the current work. 
\section*{Physical mechanism and asymptotically large violations}
The interferometer shown in Fig.~\ref{fig:opticalScheme} arguably provides only a small relative violation of the generalized bunching conjecture, as seen from Eq.~\eqref{eq:violation}.  It is natural to ask whether larger relative violations can be obtained and what would be the corresponding physical set-up. Moreover, from a physics perspective, it is important to pinpoint the underlying mechanism that explains the violation, at least in some particular setting. We answer these questions and find a way to generalize the \mbox{7-mode} circuit into an \mbox{$n$-mode} circuit, as described in the caption of Fig.~\ref{fig:opticalScheme}. For this family of circuits, we show that the ratio of the bunching probabilities, which we refer to simply as the \emph{bunching violation ratio} $R_n$, obeys the following bound 
\begin{equation}
    \label{eq:bunchingRatio}
    R_n= \frac{P_n^{({\displaystyle\star})}} {\Pbos{n}}  \geq \frac{n}{8} + \frac{1}{32}\frac{(n-2)^2}{n-1},
\end{equation}
for any $n\geq 4$. Hence, partial distinguishability can lead to an asymptotically larger multimode bunching probability with respect to fully indistinguishable bosons. While a detailed derivation of this bound is given in Appendix \ref{sec:bound}, we present here the main arguments by comparing the physical mechanism of bunching for fully indistinguishable and partially distinguishable photons.

\subsubsection*{Fully indistinguishable photons.} The first step of the argument is to note that the only possibility for all of the $n$ photons to be observed in the subset $\mathcal{K}$ is if there are $q=n-2$ photons in the first two output modes of the DFT interferometer of dimension $q$ and vacuum on the rest (we assume $q\geq 2$). The corresponding conditional (subnormalized) state of these two output modes is given by the NOON state \cite{pryde2003creation}
\begin{align}\label{eq:NOON}
  \ket{\psi_\mathrm{out}^\mathrm{(bos)}}
  &=\frac{1}{q^{q/2}}  \prod _{j=0}^{q-1} \left( \ad_0 + \omega^j\,\ad_1  \right) \ket{0} \nonumber\\  
  &=\frac{1}{q^{q/2}}
  \left((\ad_0)^{q} + (-1)^{q} (\ad_1)^{q}\right)\ket{0},
\end{align}
where $\omega = \exp \left(2 i \pi / q\right)$ is the $q$th root of unity. The probability of having $q$ photons in these two modes is simply given by the square norm of this state, namely $2q!/q^q$. The next part of the circuit realizes the interference between modes $0$ and $0'$ via beam splitter $\hat{U}_\mathrm{BS}^{0,0'}$, as well as between modes $1$ and $1'$ via beam-splitter $\hat{U}_\mathrm{BS}^{1,1'}$. The action of these beam-splitters on state  $\ad_{0'}\ad_{1'}\ket{\psi_\mathrm{out}^\mathrm{(bos)}}$ followed by postselection on vacuum in both output modes 0 and 1 is analyzed in Appendix \ref{sec:bound}. The resulting probability is governed by the bunching mechanism sketched in the upper part of  Fig.~\ref{fig:buncingBS}, where $\eta$ denotes the transmittance of the two beam splitters. The first term of state \eqref{eq:NOON} describing $q$ photons in mode 0 undergoes bunching with the extra photon in mode 0' with probability $(q+1) \,\eta(1-\eta)^q $, while the extra photon in mode 1' is simply transmitted with probability $\eta$. The second term of state \eqref{eq:NOON} behaves similarly.   Consequently, the multimode bunching probability (in output modes $0'$ and $1'$) is given by
\begin{equation}\label{eq:bunching_prob_bos}
    \Pbos{n} = \frac{2\, (q+1)!}{q^{q}} \, \eta^2 (1-\eta)^{q} 
\end{equation} 
\begin{figure}
    \centering
    \includegraphics[width = 0.4\textwidth]{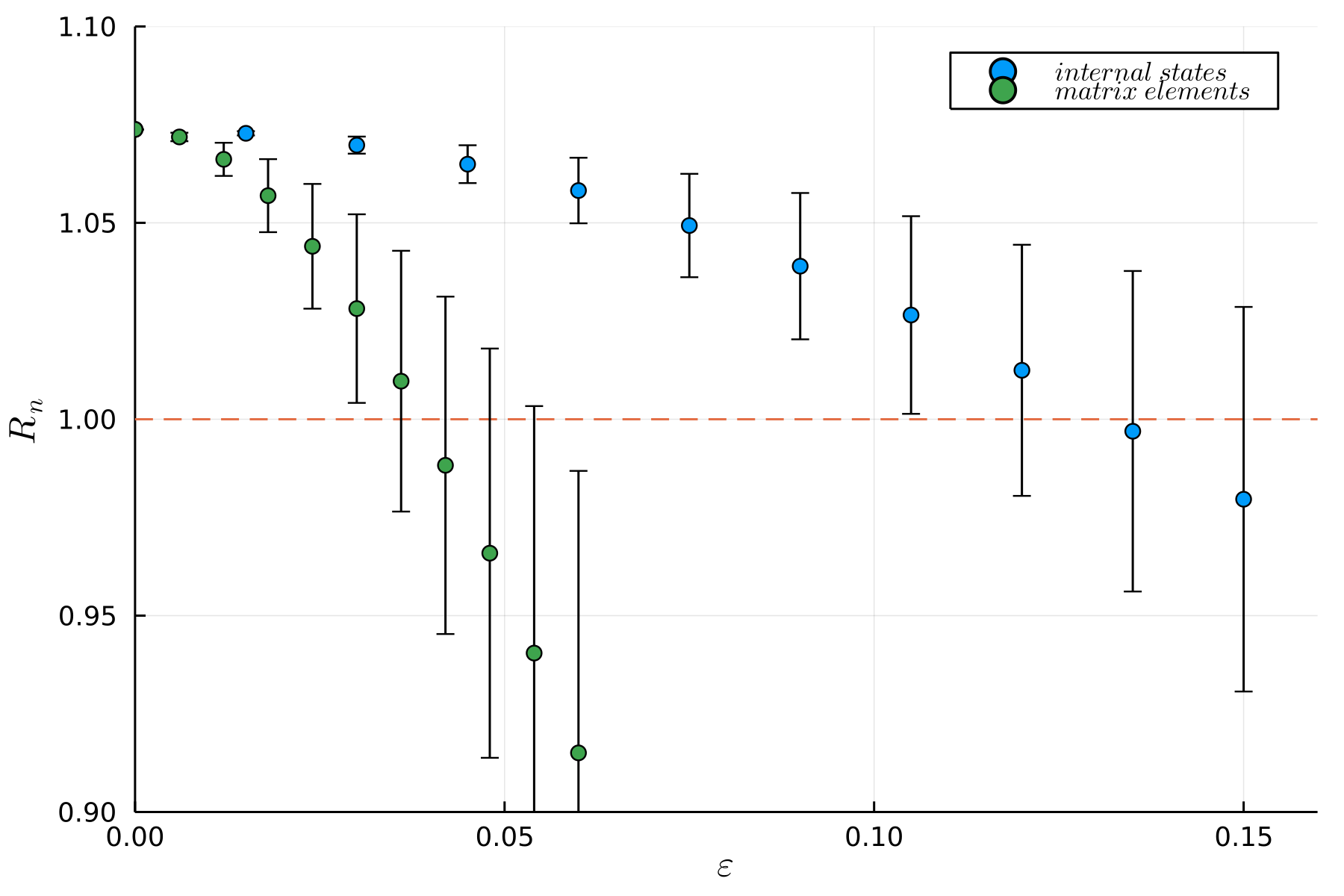}
    \caption{Perturbation effects on the bunching violation ratio $R_n$ (vertical axis) for the setup of Fig.~\ref{fig:opticalScheme} with $n=7$. (Blue) Perturbation of the internal wave functions 
    defined in the caption of Fig.~\ref{fig:opticalScheme} by a random amount drawn from a Gaussian distribution with zero mean and standard deviation $\epsilon$ (horizontal axis). For each  $\epsilon$ value, $10^4$ samples are taken. We observe violations of Conjecture \ref{conj:generalizedBunching} (on average) up to $\epsilon_{max} \approx 0.135$, which corresponds to the components of the internal wave functions being perturbed by about $\epsilon_{max}/(1+\epsilon_{max}^2) \approx13.3\%$. The vertical bars represent the standard deviation for each $\epsilon$ value.
    (Green) Perturbation of the matrix elements $U_{ij}$ of the 7-mode interferometer. The same random Gaussian perturbations are added to the columns of the matrix $U$, which are then Gram-orthonormalized. In this case, the matrix elements can be perturbed by about $\epsilon_{max} \approx 0.039$ to still exhibit a violation (on average). The optical scheme is thus resilient to perturbations both to the internal states of the photons and to the interferometer, making a good case for its experimental feasibility.  }
    \label{fig:gramPerturbations}
\end{figure}

\subsubsection*{Partially distinguishable photons.} 
\begin{figure}[t]
    \centering
    \includegraphics[width = 0.4\textwidth]{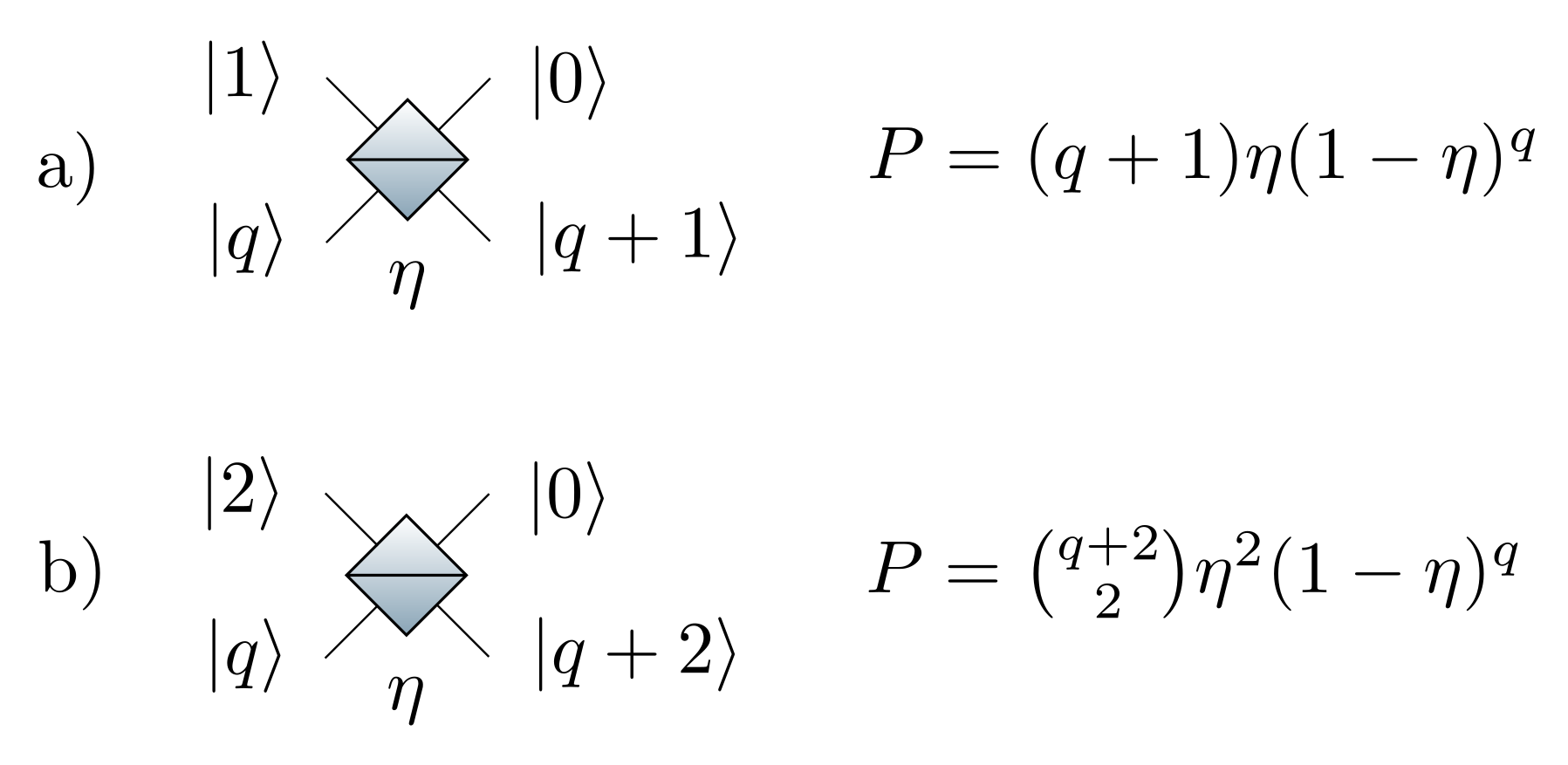}
    \caption{Mechanism at the origin of boosted bunching. (a) For indistinguishable photons, the extra photon in mode $\ad_{0'}$ (or $\ad_{1'}$) bunches with $q$ photons in mode $\ad_{0}$ (or $\ad_{1}$) coming from the NOON state~\eqref{eq:NOON}. (b) For partially distinguishable photons, the largest term contributing to the bunching probability~\eqref{eq:bunching_prob_pd} comes from the {\it double bunching} of two photons in the delocalized mode $\hat{c'}^\dagger_+$ [see Eq.~\eqref{eq:basis_change_2}] with $q$ photons in the delocalized mode $\hat{c}^\dagger_+$ [see Eq.~\eqref{eq:basis_change_1}]. The asymptotics of the bunching violation ratio $R_n$ as shown in Eq.~\eqref{eq:bunchingRatio} originates from the probabilities of the processes depicted here. }
    \label{fig:buncingBS}
\end{figure}

\begin{figure}[t]
    \centering
    \includegraphics[width = 0.4\textwidth]{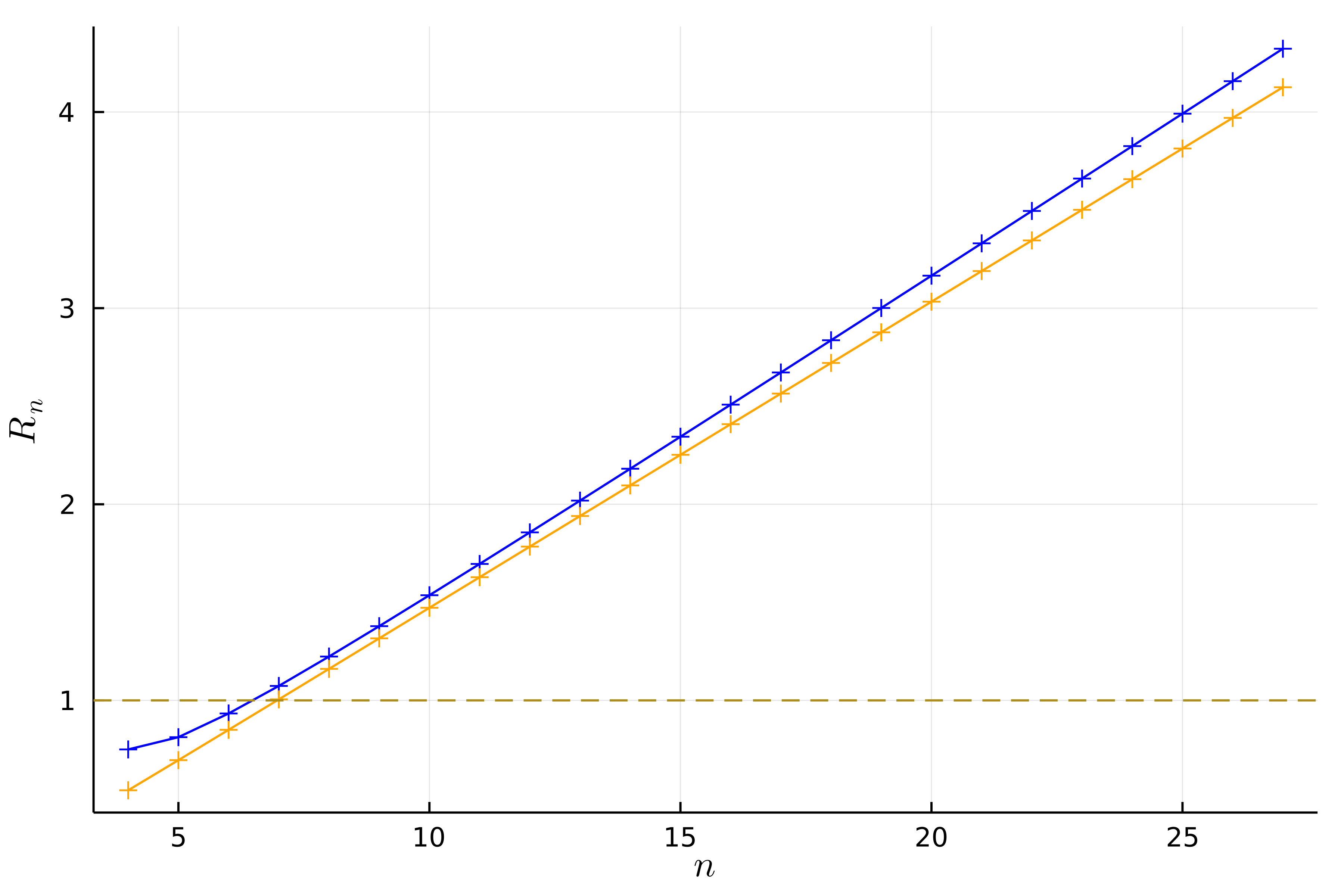}
    \caption{Bunching violation ratio $R_n$ for the family of optical schemes shown in Fig.~\ref{fig:opticalScheme} as a function of the size $n$ (blue). For $n\ge 7$, it appears that partially distinguishable particles outperform indistinguishable bosons as witnessed by $R_n >1$.  The lower bound given by Eq. \eqref{eq:bunchingRatio} is also shown (orange).}
    \label{fig:bunchingRatio}
\end{figure}

We consider the following $q$-star polarization pattern   
\begin{equation}\label{eq:psiin_pd}
    \ket{\psi_\mathrm{in}^{({\displaystyle\star})}} = \frac{1}{2^{q/2}}\prod _{j=0}^{q-1} \left(\ahd{j} + \omega^j\, \avd{j} \right)\ket{0}, 
\end{equation}
as the input state sent to the DFT interferometer. Here, $\ahd{j}$ ($\avd{j}$) are creation operators of a photon in spatial mode $j$ and horizontal (vertical) polarization. This state is a generalization of the 5-star polarization pattern shown in Fig. \ref{fig:opticalScheme} (right).  In order to understand why multimode bunching is boosted with this special input state, it is convenient to define the spatio-polarization modes
\begin{align}\label{eq:basis_change_1}
\hat{c}^{\dagger}_{\pm}&= \dfrac{\ahd{1}\pm \avd{0}}{\sqrt{2}}.
\end{align}
Similarly as for indistinguishable photons, we compute the conditional output state of the DFT interferometer that contains $q$ photons in the first two output modes (and vacuum for the rest) for input state~\eqref{eq:psiin_pd}, namely
\begin{equation}
\ket{\psi_\mathrm{out}^{({\displaystyle\star})}}=\frac{1}{q^{q/2}}\prod _{j=0}^{q-1} \left( \frac{\ahd{0}}{\sqrt{2}}+ \omega^j\,  \hat{c}^{\dagger}_+ + \omega^{2j}\,  \frac{\avd{1}}{\sqrt{2}}  \right) \ket{0} .
\end{equation}
This polarization multi-mode state 
is the counterpart of the NOON state, Eq.~\eqref{eq:NOON}. It comprises many terms but we solely focus here on the term that gives the largest asymptotic contribution to the multimode bunching probability, namely 
\begin{align}\label{eq:psiout_pd}
\ket{\psi_\mathrm{out}^{({\displaystyle\star})}}= \dfrac{(-1)^{q+1}} {q^{q/2}}(\hat{c}^{\dagger}_{+})^{q}\ket{0}+\dots
\end{align}
The other components of state~\eqref{eq:psiout_pd} are orthogonal to the state $(c^{\dagger}_+)^{q}\ket{0}$ and thus can only contribute with additional positive terms to the bunching probability. The probability to have $q$ photons (regardless of their polarization) in the first two modes is thus lower bounded by $q!/q^q$.
The subsequent part of the interferometer is fed with the state
\begin{align}\label{eq:interference_pd_input}
&\ahd{1'} \, \avd{0'}\ket{\psi_\mathrm{out}^{({\displaystyle\star})}} = \frac{(\hat{c'}^{\dagger}_{+})^2-(\hat{c'}^{\dagger}_{-})^2}{2} \ket{\psi_\mathrm{out}^{({\displaystyle\star})}}, 
\end{align}
where we have described the two extra photons with antipodal polarization (H and V) using the spatio-polarization modes
\begin{align}
\hat{c'}^{\dagger}_{\pm}&= \dfrac{\ahd{1'}\pm \avd{0'}}{\sqrt{2}}, \label{eq:basis_change_2}
\end{align} 
defined in analogy with Eq. \eqref{eq:basis_change_1}. Thus, 
the leading term of the final output state of the interferometer is
\begin{align}\label{eq:interference_pd}
 \frac{(-1)^{q+1}}{2 \, q^{q/2}} \,  \hat{V}\left((\hat{c'}^{\dagger}_{+})^2-(\hat{c'}^{\dagger}_{-})^2\right)(\hat{c}^{\dagger}_{+})^{q}\ket{0}+\dots, 
\end{align}
where we have defined $\hat{V}=\hat{U}_{BS}^{0,0'}\, \hat{U}_{BS}^{1,1'}$ as the operator describing the joint operation of the two beam-splitters in both polarization (see Appendix \ref{sec:bound}). Using the fact that the beam splitters have the same transmittance $\eta$, it appears that $\hat{V}$ also acts as a beam splitter of transmittance $\eta$ that couples modes $\hat{c}^{\dagger}_{+}$ and $\hat{c'}^{\dagger}_{+}$ \footnote{The beam splitter $\hat{V}$ also couples $\hat{c}^{\dagger}_{-}$ and $\hat{c'}^{\dagger}_{-}$ but we disregard the corresponding term here as its contribution is much smaller and we seek a lower bound on the probability, see Appendix \ref{sec:bound}.}. We thus have interference between $q$ photons in  mode $\hat{c}^{\dagger}_{+}$ (occurring with probability $q!/q^q$) and two photons in mode $\hat{c'}^{\dagger}_{+}$ (occurring with probability 1/2). Hence, the resulting probability of obtaining $n=q+2$ photons in output mode $\hat{c'}^{\dagger}_{+}$ (which leads to the detection of $n$ photons in the output bin $\mathcal{K}=\{0',1'\}$) is governed by the double-bunching mechanism sketched in the lower part of  Fig.~\ref{fig:buncingBS}, 
associated with probability ${q+2 \choose 2} \, \eta^2 (1-\eta)^q$. 
As a result, 
we obtain the following bound on the 2-mode bunching probability in the case of partially distinguishable photons
\begin{equation}\label{eq:bunching_prob_pd}
 P_n^{({\displaystyle\star})} \geq \frac{(q+2)! }{4\, q^{q}} \, \eta^2 (1-\eta)^{q} .
\end{equation}
This probability is asymptotically larger than its counterpart for fully indistinguishable photons, Eq.~\eqref{eq:bunching_prob_bos}. Using Eqs.~\eqref{eq:bunching_prob_bos} and \eqref{eq:bunching_prob_pd}, we indeed obtain a lower bound on the bunching violation ratio 
\begin{equation}
R_n = \frac{ P_n^{({\displaystyle\star})} } {\Pbos{n}} \geq \frac{q+2}{8} = \frac{n}{8} \, ,
\end{equation}
which confirms the dominant term in Eq.~\eqref{eq:bunchingRatio} and shows that it grows at least linearly with $n$. A more detailed calculation given in Appendix \ref{sec:bound} leads to the second term in Eq.~\eqref{eq:bunchingRatio}. As seen in Fig.~\ref{fig:bunchingRatio}, this bound seems to describe well enough the true behavior of $R_n$ up to $n=30$. In the special case where $n=7$, we may compute exactly all terms in Eq.~\eqref{eq:psiout_pd}, which gives $R_7=1237/1152$, in perfect agreement with Eq. \eqref{eq:violation}.  Note also that Eq.~\eqref{eq:bunchingRatio} shows no dependence on the transmittance $\eta$. However, it can be seen that the absolute probability of bunching events is maximized by maximizing the term $\eta^2 (1-\eta)^{q}$ over $\eta$, which yields $\eta = 2/n$ as mentioned in the caption of Fig.~\ref{fig:opticalScheme}.
\vspace{7pt}

%
\section*{Discussion}
The complex behavior of interferometric experiments with multiple partially distinguishable photons has been explored in several theoretical and experimental works \cite{tichy2011four, ra2013nonmonotonic, tichyTutorial,tillmann2015generalized, Turner2016PostselectiveQI, interferingDistPhotons, menssen2017distinguishability, shchesnovich2018collective,jones2019distinguishability}, revealing that many-body interference does not reduce to a simple dichotomy between distinguishable and indistinguishable photons. This is evident, for example, from the fact that certain outcome probabilities do not behave monotonically as one makes photons more distinguishable \cite{tichy2011four, ra2013nonmonotonic}. However, the scheme of Fig.~\ref{fig:opticalScheme} is the first evidence that boson bunching can be boosted by partial distinguishability to the point where it actually beats ideal (fully indistinguishable) bosons. This disproves the common belief that bunching effects are maximized in this ideal scenario.

It is intriguing to observe that the state of partially distinguishable photons we have found to exhibit boosted bunching is, in a sense, \emph{far} from the state of fully indistinguishable photons. This can be seen by computing the relative contribution of the fully (permutation-) symmetric component of the internal wavefunction $\ket{\Phi} = \ket{\phi_1}\ket{\phi_2}\dots\ket{\phi_n}$ \cite{shchesnovich2015tight}
\begin{equation}
d(S) = \braAket{\Phi}{\op{\mathcal{S}}_n }{\Phi} = \frac{\perm{S}}{n!}, 
\end{equation}
where $\op{\mathcal{S}}_n = (1/n!)\sum _{\sigma \in S_n} \op{P}_\sigma$ is the symmetrizer in $n$ dimensions.
This measure is $1$ for fully indistinguishable photons ($S = \mathbb{E}$) and $1/n!$ for fully distinguishable ones ($S=\mathbb{1}$). For the simplest case of seven photons, we have that
\begin{align}
    d(\mathbb{1}) &= \frac{1}{7!}\approx 1.98\times 10^{-4}\nonumber \\ 
  d(S^{({\displaystyle\star})}) &= \frac{45}{7!} \approx 8.93\times 10^{-3} \ll 1 
\end{align}
where $S^{({\displaystyle\star})}$ is the Gram matrix of the partially distinguishable polarization state shown in Fig.~\ref{fig:opticalScheme}. It is thus natural to ask whether there may exist states violating the generalized bunching conjecture already \emph{in the vicinity} of a fully indistinguishable state. We can show that first-order perturbations around $S = \mathbb{E}$ leave the multimode bunching probability constant, which suggests a negative answer (see Appendix~\ref{sec:stability}). But the question remains open whether this probability always decreases near $S = \mathbb{E}$ if second-order terms are taken into account, in which case it would be a general feature.

Furthermore, it must be noted that we have modeled distinguishable photons with a two-dimensional internal degree of freedom, namely polarization. In contrast, in the fully distinguishable setting, each photon occupies a different internal state, forming an orthonormal basis of an $n$-dimensional space. Is it possible to find counterexamples where the internal states live in a larger space, going beyond small perturbations around our conterexamples? This could model realistic situations with photons occupying partly overlapping time-bins. We believe this is possible and should lead to larger than linear asymptotic scalings of the bunching violation ratio. We leave this question for future work.

On a final note, we stress that our findings corroborate the deep connections between bosonic interferences in quantum physics on the one hand, and the algebra of matrix permanents on the other hand. The transposition of Drury's \mbox{7-dimensional} matrix counterexample into a quantum optical interferometric experiment has inspired us to find a family of $n$-dimensional matrices that not only violate the Bapat-Sunder conjecture but also exhibit a relative violation increasing with $n$. We anticipate that other mathematical conjectures on permanents may be addressed by exploiting this fruitful interplay with physics-inspired mechanisms such as those shown in Fig.~\ref{fig:buncingBS}. This may even help solve other questions on the Bapat-Sunder conjecture \cite{Zhang+2016}. For example, the smallest known counterexample is $7$-dimensional, so it would be interesting to find a simpler counterexample if it exists, or show that this is not possible. Another open question in matrix theory is whether there is a counterexample involving a real matrix of dimension smaller than 16 \cite{drury2017real}, which may be resolved by considering interferometry within real quantum mechanics.

Overall, we hope this work will open new paths to explore the connection between distinguishability and boson bunching, leading not only to a better understanding of multiparticle quantum interference but perhaps also to novel applications of partially distinguishable photons to quantum technology. 
\section*{Acknowledgments} 
The authors would like to thank S. Drury for useful correspondence as well as F. Flamini and V. \valery~for valuable discussions. B.S. and L.N. are respectively a Research Fellow and Postdoctoral Researcher of the Fonds de la Recherche Scientifique – FNRS. N.J.C. acknowledges  support  by  the  Fonds de la Recherche Scientifique – FNRS  under  Grant No T.0224.18  and  by  the  European Union  under project ShoQC within ERA-NET Cofund in Quantum Technologies (QuantERA) program. This project has also received funding from the European Union’s Horizon 2020 research and innovation programme under the Marie Skłodowska-Curie grant agreement No 956071. Finally, we thank Alexander Franzen for ComponentLibrary used in making the figures. 

\appendix
\section*{Appendix}


\section{Bunching probability}

\label{sec:bunchingProbability}
In this section we summarize the main steps needed to derive Eq. \eqref{eq:bunchingProbability}. Following the colloquial conventions of \cite{tichy2010zero}, we consider $n$ photons sent through a $(m,m)$ linear interferometer described by the unitary matrix $U$.
We limit ourselves to at most one photon per input mode.
Without loss of generality, we consider that the photons occupy the first $n$ input modes. 
We denote the vector of occupation number of the output modes as $\vect{s} = (s_i)$ where $0 \leq s_i \leq n$ is the number of photons in output mode $i$. Naturally $\sum s_i = n$. We define the mode assignment list $\vect{d} = \vect{d}(\vect{s}) = \oplus _{j = 1}^n \oplus _{k = 1}^{s_j} (j)$. For instance if $\vect{s} = (2,0,1)$ then $\vect{d} = (1,1,3)$.

Consider the probability $P(\vect{d})$ that the photons give an output configuration $\vect{s}$ with a  mode assignment list $\vect{d} = \vect{d}(\vect{s})$. Tichy shows that this probability can be expanded as a multi-dimensional tensor permanent \cite{tichy2015_partial_distinguishability}
\begin{equation}
    P(\vect{d}) =\frac{1}{\mu(\vect{s})} \sum_{\sigma, \rho \in S_{n}} \prod_{j=1}^{n}\left(U_{\sigma_{j}, d_j} U_{\rho_{j}, d_j}^{*} S_{\rho_j, \sigma_{j} }\right) 
\end{equation}
with $\mu(\vect{s}) = \prod _{j=1}^m s_j!$.

Let us now compute the probability $P_n(S) $ that all $n$ photons bunch in a subset $\mathcal{K}$ of the output modes, for a set interferometer and a Gram matrix $S$, following a derivation of Shchesnovich \cite{shchesnovich2016universality}. 
Without loss of generality, consider that the subset $\mathcal{K}$ is the first $K = |\mathcal{K}|$ output modes. 
The bunching probability is the sum over all event probabilities $P(\vect{d})$ with $s_i = 0$ for all $i > K$
\begin{equation}
    P_n(S) = \frac{1}{n!} \sum _{d_1 = 1}^K \dots \sum _{d_n = 1}^K \sum_{\sigma, \rho \in S_{n}} \prod_{j=1}^{n}\left(U_{\sigma_{j}, d_j} U_{\rho_{j}, d_j}^{*} S_{\rho_{j},\sigma_{j}}\right) 
\end{equation}
Now, calling 
\begin{equation}
    H_{a,b} = \sum _{i = 1} ^K U_{i,a}^* U_{i,b} 
\end{equation}
we can rewrite
\begin{equation}
    P_n(S) =  \sum_{\sigma'\in S_{n}} \prod_{j=1}^{n}\left(H_{j, \sigma'_j} \Stranspose_{j, \sigma'_{j}}\right) = \perm{H \odot \Stranspose}
\end{equation}
where the last quantity is the permanent of the Hadamard (or elementwise) product $(H\odot \Stranspose)_{ij} \equiv H_{ij}S_{ji}$. Note that for indistinguishable particles $S_{ij} = 1, ~\forall i,j$, so that $P_n(\mathbb{E}) = \perm{H}.$

\section{Physical realization of violating matrices}
\label{sec:physicalRealization}
It is possible to see that any counterexample to the Bapat-Sunder conjecture can be used to construct a physical interferometer $U$ and a set of internal states of the photons $\{\ket{\phi_i}\}$ that provide a counterexample to the generalized bunching conjecture. We assume, without loss of generality \cite{zhang1989notes}, a simplified form of the Bapat-Sunder conjecture, where $A$ and $B$ are Gram matrices and so $a_{ii}=b_{ii}=1, \forall i\in\{1,...,n\}$. In this case,  Conjecture \ref{conj:bapatSunder} takes the form $\perm{A\odot B} \leq \perm{A}$.  
Since the distinguishability matrix $S$ is a Gram matrix, we can choose $\Stranspose=B$. The set of quantum states realizing any given distinguishability matrix can be obtained from its Cholesky decomposition 
\begin{equation}
\label{eq:cholesky}
    B = M'^\dagger M'.
\end{equation}
The matrix $M'$ is of size $r'\times n$ where $r'$ is the rank of $B$. The $n$ internal photon states $\{\ket{\phi_i}\}$ that realize this Gram matrix can be read out from the columns of $M'$. Thus, the rank of $B$ determines the dimension of the Hilbert space spanned by the states $\{\ket{\phi_i}\}$. For the physical realization of Drury's counterexample we chose $S=A=M^\dagger M$, with $M$ given in Eq.~\eqref{eq:internalWaveFunctions}. This implies that the  7 internal states of the bosons live in a two-dimensional space, where each state is obtained from each column of $M$ .   

In addition, it is always possible to construct an interferometer $U$ such that $H= \alpha A$, where $\alpha$ is a positive rescaling factor such that $\alpha\leq 1$. Note that this rescaling is not important when it comes to showing a violation of the generalized bunching conjecture~\footnote{If $\perm{A\odot B}\geq \perm{A}$ then $\perm{\alpha A\odot B}\geq \perm{\alpha A}$}.  
We can write the Cholesky decomposition of $\alpha A$ as 
\begin{equation}
H_{a,b} = \alpha \sum _{k=1}^r M_{a,k}^\dagger M_{k,b} = \alpha \sum _{k=1}^r M_{a,k}^\dagger \left(M_{b,k}^\dagger\right)^*, \end{equation}
where $r$ is the rank of $A$. By comparing with the definition of the matrix $H$ in Eq.~\eqref{eq:H_Shchesnovich}, it is possible to see that we obtain $H=\alpha A$ if we appropriately incorporate the matrix $\sqrt{\alpha} M^{\dagger} $ as a submatrix of $U$, for example, in the top left corner. This choice determines that the subset $\mathcal{K}$ is given by the first $r$ output modes. Note also that it is always possible to incorporate an arbitrary complex matrix, up to renormalization, into a bigger unitary matrix, using arguments similar to Lemma 29 of Ref.~\cite{aaronson2011computational}.

In the case discussed in the main text, the aim is to construct an interferometer $U$ which contains a rescaled version of the $7\times 2$ matrix $\sqrt{\alpha} M^{\dagger}$, where $M$ is given in Eq.~\eqref{eq:internalWaveFunctions}. Here, the procedure to construct $U$ is simplified by the fact that the columns of $M^\dagger$ are already orthogonal vectors. Hence, we can choose $\alpha=2/7$ to normalize these columns and find 5 other orthonormal vectors to construct a $7\times 7$ unitary matrix. The unitary built from the circuit presented in Fig.~\ref{fig:opticalScheme} gives one possibility to construct such a unitary, which was chosen for its simplicity.  
%
\section{Resilience to perturbations} 
\label{sec:resilience}
It is natural to ask whether small perturbations of the interferometer $U$ or the partial distinguishability matrix $S$ would immediately negate the violation of generalized bunching conjecture, or, equivalently of the Bapat-Sunder's conjecture, in which case the discussion would be physically insignificant. However it is easy to see that this is not the case as we now show. 

We consider for instance a perturbation of the matrix $A$ in $\perm{A\odot B} > \perm{A}$, the reasoning being similar if we also consider perturbations in the matrix $B$. We consider the perturbation $A \rightarrow (1-\epsilon)A + \epsilon \Delta$ with $\Delta_{i,i} = 1, \Delta_{i,j} = \mathcal{O}(1)$, with $\epsilon\ll 1$. Note that 
\begin{align}
    &\perm{(1-\epsilon)A + \epsilon \Delta} \nonumber\\
    &= \perm{(1-\epsilon)\left(A + \frac{\epsilon}{(1-\epsilon)} \Delta\right)} \nonumber\\
     &= (1-\epsilon)^n\perm{A + \delta \Delta}
\end{align}
setting $\delta = \frac{\epsilon}{(1-\epsilon)}  = \mathcal{O}(\epsilon)$. The same factorization may be operated on both sides of the Bapat-Sunder inequality.

Next, we present a formula from Minc \cite{minc1984permanents} for the permanent of the a sum of two matrices 
\begin{align}
    &\perm{A+B} = \nonumber\\
    &\sum_{r=0}^n \sum_{\alpha,\beta \in Q_{r,n}} \perm{A[\alpha,\beta]}\perm{B(\alpha,\beta)}
\end{align}
where $Q_{r,n}$ is the set of increasing sequences. More precisely, if we denote $\Gamma_{r, n}$ as the set of all $n^{r}$ sequences $\omega=\left(\omega_{1}, \ldots, \omega_{r}\right)$ of integers, $1 \leq\omega_{i} \leq n, i=1, \ldots, n$, we define 
\begin{equation}
Q_{r, n}=\left\{\left(\omega_{1}, \ldots, \omega_{r}\right) \in \Gamma_{r, n} \mid 1 \leq \omega_{1}<\cdots<\omega_{r} \leq n\right\}.
\end{equation}
Moreover, $A[\alpha,\beta]$ denotes the $r\times r$ matrix constructed by choosing the rows and columns of $A$ according to the sets $\alpha$ and $\beta$. In contrast, $A(\alpha,\beta)$ is the $(n-r)\times (n-r)$ matrix where those rows and columns have been excluded. 
This way, we can expand to first order 
\begin{align}
&\perm{A + \delta \Delta} =\nonumber\\& =\perm{A} + \delta \sum _{i,j = 1}^n \Delta_{i,j}\perm{A(i,j)}+ \mathcal{O}(\delta^2)\nonumber\\ 
     &\perm{(A + \delta \Delta)\odot B} =\nonumber\\
     &=\perm{A\odot B}+ \delta \sum _{i,j = 1}^n \Delta_{i,j}B_{i,j}\perm{(A\odot B)(i,j)}\nonumber\\&~~~~ + \mathcal{O}(\delta^2)
\end{align}
so that if $\perm{A\odot B} > \perm{A}$ then, for small enough $\epsilon$, we have that
\begin{equation}
\perm{((1-\epsilon)A + \epsilon \Delta)\odot B} > \perm{(1-\epsilon)A + \epsilon \Delta},
\end{equation}
meaning that small enough perturbations to the matrix $A$ still lead to a violation of the Bapat-Sunder inequality. A similar reasoning can be used to argue about robustness to perturbations to matrix $B$. For the particular counterexample of the generalized bunching conjecture considered in the main text, the robustness to perturbations can be seen in Fig.~\ref{fig:gramPerturbations}. 

\begin{figure}
    \centering
    \includegraphics[width = 0.4\textwidth]{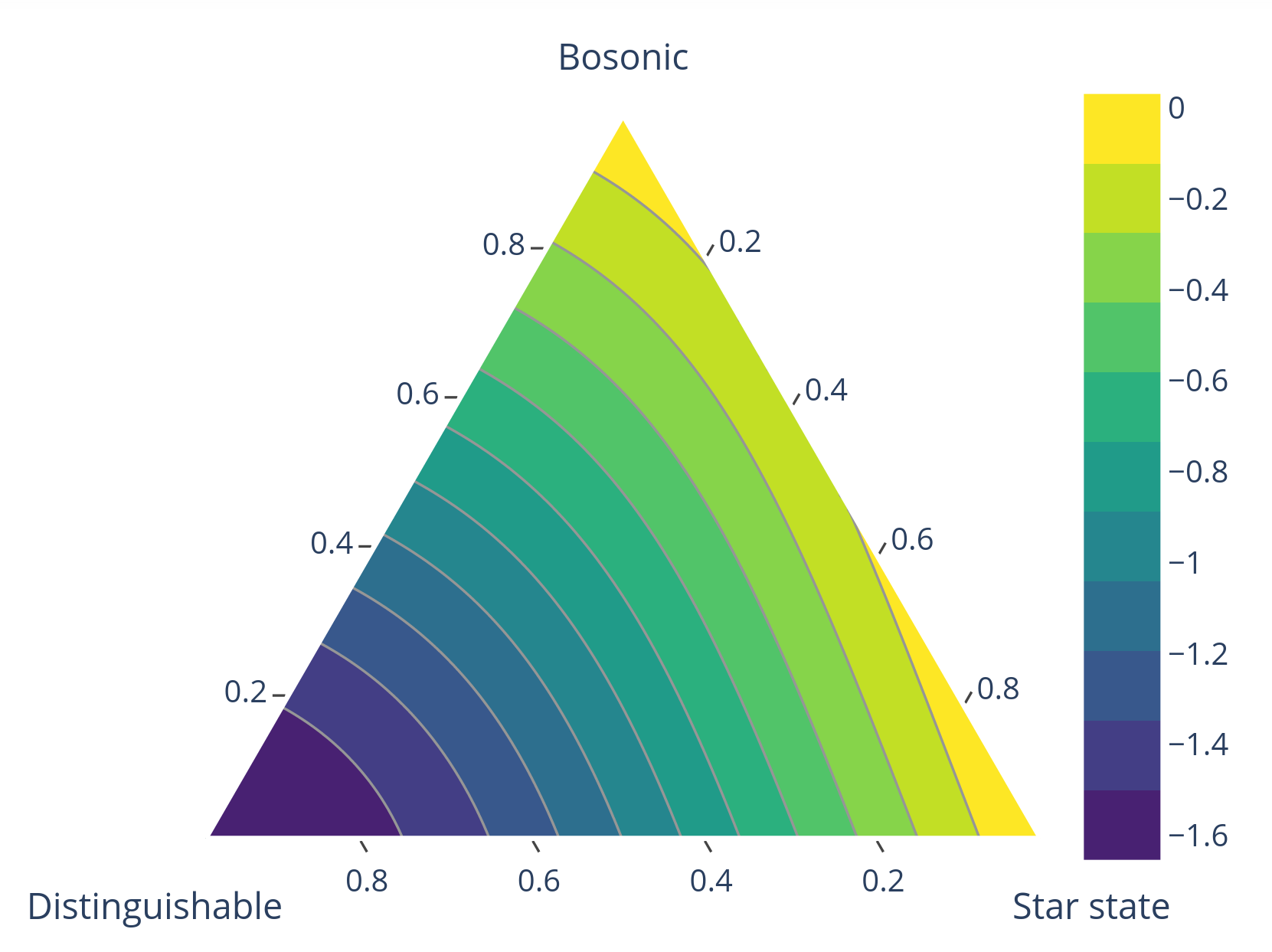}
    \caption{Ternary plot of the $\log_{10}$ of the bunching ratio $P_7(S(x,y))/\Pbos{7}$ for the optical scheme from Fig.~\ref{fig:opticalScheme}, and the two-parameter family of distinguishability matrices $S(x,y)$ defined in Eq.~\eqref{eq:S_interpol}. }
    \label{fig:ternary_plot}
\end{figure}

In order to further visualize how the choice of distinguishability matrix can affect the bunching violation ratio, we consider the following two-parameter family of $S$ matrices of dimension $7$
\begin{equation}\label{eq:S_interpol}
S(x,y) = (1-x-y)S^{({\displaystyle\star})}+x S^{\mathrm{(bos.)}} + y,  S^{\mathrm{(dist.)}}. 
\end{equation}
where $x,y\geq0$ and $x+y\leq 1$. Here, $S_{\displaystyle\star}$ corresponds to the $S$ matrix of the partially distinguishable input state from Fig.~\ref{fig:opticalScheme}, whereas $S_{\mathrm{bos}}=\mathbb{E}$ and $S_{\mathrm{dist}}=\mathbb{1}$ correspond to the fully indistinguishable and fully distinguishable cases, respectively. The bunching violation ratio for these different $S$ matrices is plotted in Fig.~\ref{fig:ternary_plot}. As one gets gets closer to the case of distinguishable particles, the bunching decreases significantly as expected. However, when we interpolate between the $S^{\mathrm{(bos.)}}$ and $S_{\displaystyle\star}$, the bunching probability behaves non-monotonically and the bunching violation ratio $P_7(S(x,y))/\Pbos{7}$ attains values larger than 1 in a small region around $S_{\displaystyle\star}$.   

%
\section{Bound on the bunching violation ratio}
\label{sec:bound}
In this section we give the detailed derivation leading to the bound on the bunching violation ratio of Eq.~\ref{eq:bunchingRatio}. This also allows for a better physical understanding of reason behind the enhanced bunching using partially distinguishable photons. 

We consider the circuit described in the caption Fig.~\ref{fig:opticalScheme}, which is a generalization of the 7-mode optical circuit depicted in this figure to a circuit of $n$ modes. The circuit is composed by a discrete Fourier transform (DFT) circuit of size $q=n-2$ applied to the input modes $\{0, 1,..., q-1\}$ followed by two beam splitters of equal transmittance applied between the modes $0'$ and $0$ as well as between the modes $1'$ and $1$.

Our quantity of interest is the probability of observing all the $n$ photons in the output modes $0'$ and $1'$. Let us first compute this quantity when the input photons are fully indistinguishable. The quantum state at the output of the DFT is given by
\begin{align}
    \hat{U}_{dft}\ket{\psi_{in}} &= \hat{U}_{dft} \prod _{j=0}^{q-1} \ad_j \ket{0} \\
    &= \frac{1}{q^{q/2}}\prod _{j=0}^{q-1}\left( \sum_{k=0}^{q-1} e^{2\pi i jk/q} \ad_k\right) \ket{0}. 
 \end{align}
The only possibility for all of the $n$ photons to be observed in modes $0'$ or $1'$ at the output of the full circuit is if at the output of the DFT interferometer there are $q=n-2$ photons in modes $0$ or $1$ and vacuum on the rest. Hence, we only consider the subnormalized component of the wavefunction in these output modes, given by 
\begin{equation}
    \ket{\psi_{\mathrm{out}}^{\mathrm{(bos)}}} = \frac{1}{q^{q/2}}\prod _{j=0}^{q-1}\left(\ad_0 + e^{2\pi i j/q} \ad_1\right) \ket{0} 
 \end{equation}
To expand this expression, one can think of $\ad_0$ and $\ad_1$ as complex numbers since these two operators commute. In this sense, following \cite{pryde2003creation}, we can consider that each term $\ad_0 + e^{2\pi i j/q} \ad_1$ is an eigenvalue of the circulant matrix of dimension $q$ given by $\operatorname{circ}(\ad_0, \ad_1, 0,\dots,0)$. Hence, the previous equation can be rewritten as 
\begin{align}
    \ket{\psi_{\mathrm{out}}^{\mathrm{(bos)}}} &= \frac{1}{q^{q/2}}  \det(\operatorname{circ}(\ad_0, \ad_1, 0,\dots,0)) \ket{0}\\
    &= \frac{1}{q^{q/2}}\left((\ad_0)^q + (-1)^q (\ad_1)^q\right)\ket{0}\label{eq:NOON_appendix},
\end{align}
which is a NOON state. The subsequent part of the interferometer couples this state with the ancillary modes $0'$ and $1'$, each containing a single photon. We denote these beam-splitters by $\hat{U}_{BS}^{0,0'}$ and $\hat{U}_{BS}^{1,1'}$ and their transmittance by $\eta$. We use the following convention for the unitary representing the action of the beam-splitter 
\begin{equation}
    U_{BS} 
    = \begin{pmatrix}
     \sqrt{\eta} & \sqrt{1-\eta} \\ 
     -\sqrt{1-\eta} & \sqrt{\eta} \\
\end{pmatrix}.
\end{equation}
The joint application of $\hat{U}_{BS}^{0,0'}$ and $\hat{U}_{BS}^{1,1'}$ results in the state
\begin{align}
&\hat{U}_{BS}^{0,0'}\hat{U}_{BS}^{1,1'}\ad_{0'}\ad_{1'}\ket{\psi_\mathrm{out}^\mathrm{(bos)}}\\
=&\dfrac{\hat{U}_{BS}^{0,0'}(\ad_0)^q\ad_{0'}\hat{U}_{BS}^{1,1'}\ad_{1'} + (-1)^q \hat{U}_{BS}^{1,1'} (\ad_1)^q\ad_{1'}\hat{U}_{BS}^{0,0'}\ad_{0'}}{q^{q/2}}\ket{0}\label{eq:output_state_bos}.
\end{align}
The postselection on the component where all photons occupy output modes $\{0',1'\}$ yields 
\begin{equation}
   \ket{\psi_{\mathrm{post}}^{\mathrm{(bos)}}}=\dfrac{ (1-\eta)^{q/2} \eta} {q^{q/2}} \left( (\ad_{0'})^{q+1} \ad_{1'} +(-1)^q (\ad_{1'})^{q+1} \ad_{0'}\right)\ket{0}
\end{equation}
Finally, the bunching probability in modes $0'$ and $1'$ is given by \begin{align}\label{eq:pn_bos_app}
\Pbos{n}&=\braket{\psi_{\mathrm{post}}^{\mathrm{(bos)}}}{\psi_{\mathrm{post}}^{\mathrm{(bos)}}}=\frac{2\, (q+1)!}{q^{q}} \, (1-\eta)^{q} \eta^2.
\end{align}

Consider now the analogous calculation for the specially chosen state of partially distinguishable photons, described in the caption of Fig.~\ref{fig:opticalScheme}. In this case, as discussed in the main text, the counterpart of the NOON state obtained in Eq.~\eqref{eq:NOON_appendix} is given by  
\begin{equation}
\ket{\psi_\mathrm{out}^{({\displaystyle\star})}}=\frac{1}{q^{q/2}}\prod _{j=0}^{q-1} \left( \frac{\ahd{0}}{\sqrt{2}}+ \omega^j\,  \hat{c}^{\dagger}_+ + \omega^{2j}\,  \frac{\avd{1}}{\sqrt{2}}  \right) \ket{0} .
\end{equation}
After the Fourier interferometer, one ancillary photon is introduced in mode $0'$ with vertical polarization and another one in mode $1'$ with horizontal polarization. At this point, the state of the system is given by 
\begin{align}\label{eq:interference_pd_input_appendix}
&\ahd{1'} \, \avd{0'}\ket{\psi_\mathrm{out}^{({\displaystyle\star})}} = \frac{(\hat{c'}^{\dagger}_{+})^2-(\hat{c'}^{\dagger}_{-})^2}{2} \ket{\psi_\mathrm{out}^{({\displaystyle\star})}}, 
\end{align}
with $\hat{c'}^{\dagger}_{\pm}$ defined in Eq.~\eqref{eq:basis_change_2}. To analyse the action of the subsequent part of the interferometer, it is useful to define the joint action of the beam-splitter operators $\hat{U}_{BS}^{0,0'}$ and $\hat{U}_{BS}^{1,1'}$ as 
\begin{equation}
    \hat{V}=\hat{U}_{BS}^{0,0'} \hat{U}_{BS}^{1,1'}. 
\end{equation}
Using the fact that both beam-splitters have equal transmittance, it can be seen that the action of $\hat{V}$ on the delocalized modes $\hat{c}^{\dagger}_{\pm}$ and $\hat{c'}^{\dagger}_{\pm}$ is given by 
\begin{align}
\hat{V}\hat{c'}^{\dagger}_{\pm}\hat{V}^{\dagger}&= \sqrt{\eta} \hat{c'}^{\dagger}_{\pm}+\sqrt{1-\eta} \hat{c}^{\dagger}_{\pm}, \\
\hat{V}\hat{c}^{\dagger}_{\pm}\hat{V}^{\dagger}&= \sqrt{\eta} \hat{c}^{\dagger}_{\pm}-\sqrt{1-\eta} \hat{c'}^{\dagger}_{\pm}. 
\end{align}
We will see that the interference between the modes $\hat{c}^{\dagger}_{+}$, which are occupied in the state $\ket{\psi_\mathrm{out}^{({\displaystyle\star})}}$, with the two bosons in mode  $\hat{c'}^{\dagger}_{+}$ from the ancillary photon state will lead to bosonic bunching effects that are responsible for the largest asymptotic contributions to the bunching probability in modes $\{0',1'\}$. In contrast, the other modes occupied in state $\ket{\psi_\mathrm{out}^{({\displaystyle\star})}}$, i.e. $\ahd{0}$ and $\avd{1}$, do not undergo any enhanced bunching effects since they do not couple neither to $\hat{c'}^{\dagger}_{+}$ nor to $\hat{c'}^{\dagger}_{-}$. For completeness, we also write the action of $\hat{V}$ on these modes
\begin{align}
\hat{V}\hat{a}^{\dagger}_{h, 0}\hat{V}^{\dagger}&= \sqrt{\eta} \hat{a}^{\dagger}_{h, 0}+\sqrt{1-\eta} \hat{a}^{\dagger}_{h, 0'}, \\
\hat{V}\hat{a}^{\dagger}_{v, 1}\hat{V}^{\dagger}&= \sqrt{\eta} \hat{a}^{\dagger}_{v, 1}+\sqrt{1-\eta} \hat{a}^{\dagger}_{v, 1'}.
\end{align}
We are now ready to analyse the action of $\hat{V}$ on the state given in Eq.~\eqref{eq:interference_pd_input_appendix}, with the aim of computing a bound for the bunching probability. For the aforementioned reasons, it will be useful to expand the state $\ket{\psi_\mathrm{out}^{({\displaystyle\star})}}$ as a superposition of states with different occupation numbers in mode $\hat{c}^{\dagger}_+$. To do so, it is useful to define 
\begin{equation}
    \hat{B}^{\dagger}_j= \frac{1}{\sqrt{2}}(\omega^{-j}\ahd{0}+\omega^{j}\avd{1}).
\end{equation}
With this definition, we can write 
\begin{equation}
\ket{\psi_\mathrm{out}^{({\displaystyle\star})}}=
\frac{(-1)^{q-1}}{q^{q/2}}\prod _{j=0}^{q-1} 
\left(\hat{c}^{\dagger}_{+}+  \hat{B}^{\dagger}_j  \right) \ket{0}.
\end{equation}
Since all operators involved in this expression commute with each other, we can expand it as if $\hat{c}^{\dagger}_{+}$ and $\hat{B}^{\dagger}_j$ were complex numbers. Precisely, we have the following expansion  
\begin{align}\label{eq:expansion}
    \prod _{j=0}^{q-1} 
\left(\hat{c}^{\dagger}_{+}+  \hat{B}^{\dagger}_j  \right)= \sum_{k=0}^q (\hat{c}^{\dagger}_{+})^{q-k} \hat{e}_k(\hat{B}^{\dagger}_0, ...,\hat{B}^{\dagger}_{q-1}),
\end{align}
where we have defined 
\begin{align}
\hat{e}_0(\hat{B}^{\dagger}_0, ...,\hat{B}^{\dagger}_{q-1})&=1,\\
    \hat{e}_k(\hat{B}^{\dagger}_0, ...,\hat{B}^{\dagger}_{q-1})&= \sum_{0\leq j_1< ...< j_k\leq q-1} \hat{B}^{\dagger}_{j_1}...\hat{B}^{\dagger}_{j_k}.
\end{align}
To simplify the notation we denote $ \hat{e}_k(\hat{B}^{\dagger}_0, ...,\hat{B}^{\dagger}_{q-1})$ simply as $\hat{e}_k$. Newton's identities give us the following recursion relation 
    \begin{equation}\label{eq:recursion}
   k \hat{e}_k =\sum_{i=1}^k(-1)^{i-1}\hat{e}_{k-i} ~\hat{p}_i, \end{equation}
    where $\hat{p}_k$ is the $k$-th power sum 
    \begin{equation}
        \hat{p}_k=\sum_{l=0}^{q-1} (\Bd_l)^k.
    \end{equation}
   These sums take a simple form for any $k\geq 1$, with
   \begin{equation}\label{eq:power_sum}
     \hat{p}_k= \begin{cases}0~~~\mathrm{for}~k~\mathrm{odd},\vspace{7pt}\\ \dfrac{q}{2^{k/2}}\binom{k}{k/2}(\ahd{0}\avd{1})^{k/2},~~~\mathrm{for}~k~\mathrm{even}. 
     \end{cases}
   \end{equation}
Using this expression for $\hat{p}_k$ together with Eq.~\eqref{eq:recursion}, it can be seen that all the terms with odd $k$ in the expansion given in Eq.~\eqref{eq:expansion} are suppressed. Moreover, these equations provide a simple way to calculate the first few terms of the expansion in Eq.~\eqref{eq:expansion} and obtain 
   \begin{equation}
      \ket{\psi_\mathrm{out}^{({\displaystyle\star})}}=    \frac{(-1)^{q-1}}{q^{q/2}} \left((\cd_+)^q-\frac{1}{2} q (\cd_+)^{q-2}\ahd{0}\avd{1}+...\right)\ket{0}. 
   \end{equation}
 The other terms of the expansion are orthogonal to the first two terms and can only contribute with additional positive terms to the bunching probability. In fact, these two terms are enough to obtain the lower bound for the bunching violation ratio presented in the main text (Eq.~\eqref{eq:bunchingRatio}). After the action of the interferometer $\hat{V}$ it can be shown that component of the wavefunction containing all the $n$ photons in modes $0'$ and $1'$ is given by 
 \begin{align}
     &\ket{\psi_\mathrm{post}^{({\displaystyle\star})}}= \nonumber \\&=\dfrac{(1-\eta)^{q/2}\eta}{2q^{q/2}} \left((\cdprime_+)^{q+2}-\frac{1}{2} q (\cdprime_+)^{q}\ahd{0'}\avd{1'} +...\right)\ket{0}. 
 \end{align}
 The omitted terms in the previous equation are orthogonal to the  first two terms. Hence, we obtain the following lower bound for the bunching probability 
 \begin{equation}\label{eq:bound_bunch_pd}
      P_n^{({\displaystyle\star})} =\braket{\psi_\mathrm{post}^{({\displaystyle\star})}}{\psi_\mathrm{post}^{({\displaystyle\star})}} \geq \frac{\eta^2 (1-\eta)^{q}}{4\, q^{q}}\left((q+2)! +\frac{1}{4}q^2 q! \right)\,  .
 \end{equation}
 Finally, we can use Eqs.~\eqref{eq:bound_bunch_pd} and Eq.~\eqref{eq:pn_bos_app} to obtain 
\begin{align}
    R_n= \frac{P_n^{({\displaystyle\star})}} {\Pbos{n}} &\geq \frac{q+2}{8} + \frac{1}{32}\frac{q^2}{q+1}\\
    &\geq \frac{n}{8} + \frac{1}{32}\frac{(n-2)^2}{n-1},
\end{align}     
 demonstrating the bound on the bunching violation ratio given in Eq.~\eqref{eq:bunchingRatio}.

\section{Stability around the bosonic case}\label{sec:stability}
In this section, we demonstrate that first order perturbations to the internal wave functions of the photons around the fully indistinguishable case leave multimode bunching probabilities invariant. We start with the internal functions of the photons equal
\begin{equation}
    \label{eq:bosonicwavefunction}
    \ket{\phi _i} = \ket{\phi _0} 
\end{equation}
for all $i = 1,...,n$. 
Let us first consider a perturbation only to the first photon's internal wavefunction
\begin{equation}
    \ket{\phi '_1} = \ket{\phi _1 + \delta \phi _1 }
\end{equation}
In order for this state to be normalized, we need
\begin{equation}
    \braket{\phi _1}{\delta \phi _1} = -i x
\end{equation}
with $x \in \mathbb{R}$ as our (small) perturbation parameter. Consider a perturbation around the bosonic case: $ S' = \mathbb{E} + \delta S $ with
\begin{equation}
     \delta S= 
 \begin{pmatrix}
  0 & -ix & \cdots & -ix \\
  ix & 0 & \cdots & 0 \\
  \vdots  & \vdots  & \ddots & \vdots  \\
  ix & 0 & \cdots & 0
 \end{pmatrix}
\end{equation}
Note that the perturbation part of the $S$ matrix is not a Gram matrix. Then 
\begin{align}
    &\perm{A \odot \Stranspose} = \perm{A + A \odot \delta \Stranspose} \nonumber \\
    &= \perm{A} + \sum _{i,j = 1}^n \delta S_{j,i}A_{i,j}\perm{A(i,j)} + \mathcal{O}(x^2)
\end{align}
To abbreviate the notations, let's call 
\begin{equation}
    F_{i,j} = A_{i,j}\perm{A(i,j)}
\end{equation}
Then the perturbation to the bunching probability is 
\begin{equation}
    \sum _{i,j = 1}^n (\delta \Stranspose \odot F)_{i,j} =  2ix \sum _{j = 2}^n \mathcal{I}(F_{1,j}) = 0
\end{equation}
where $\mathcal{I}$ stands for the imaginary part, the equation standing as
$\sum _{j = 1}^n F_{1,j}= F_{1,1} + \sum _{j = 2}^n F_{1,j} = \perm{A} \geq 0$ and $F_{1,1} \geq 0$ as $A$ is psd. Thus 
\begin{equation}
    \perm{A \odot ( \mathbb{E} + \delta \Stranspose)} = \perm{A} + \mathcal{O}(x^2)
\end{equation}
The same procedure can be applied when each photon is modified by a small quantity $\braket{\phi _0}{\delta \phi _i} =-i  x_i = \mathcal{O}(x)$ with the same result to second order.
This means that a small physical perturbation of the internal wavefunctions of the photons around the bosonic case cannot lead to a violation (and in fact leads to an identical bunching probability at first order). We leave open the question of whether the case of fully indistinguishable particles corresponds to a local maximum of any multimode bunching probability. This would require showing that second order perturbations always have a negative contribution to this probability.

\section*{Code availability}
\noindent The following project relies on the packages \textsc{Permanents.jl} and \textsc{BosonSampling.jl}.

 

\clearpage
\bibliography{references}
\bibliographystyle{unsrt}

\end{document}